\documentstyle[epsfig,a4p,12pt,rotating]{article}
\begin{document}
%
%
\newcommand{\PPEnum}    {CERN-EP/98-194}
\newcommand{\PNnum}     {OPAL Physics Note PN-306}
\newcommand{\TNnum}     {OPAL Technical Note TN-xxx}
\newcommand{\Date}      {November 23, 1998}
\newcommand{\Author}    {Chung~Yun~Chang}
\newcommand{\MailAddr}  {Chang@umdhep.umd.edu}
\newcommand{\EdBoard}   {Peter Clarke, Paolo Giacomelli,\\ 
        George Snow and Satoru Yamashita}
\newcommand{\DraftVer}  {Version 2.0}
\newcommand{\DraftDate} {\today}
\newcommand{\TimeLimit} { 9:00 am on September 20, 1998 }

\def\toprule{\noalign{\hrule \medskip}}
\def\midrule{\noalign{\medskip\hrule }}
\def\botrule{\noalign{\medskip\hrule }}
\setlength{\parskip}{\medskipamount}


\def\dat#1{\hskip 3.3in {#1}\\}
\def\Z{\rm{Z}^{0}}
\def\Zs{\rm{Z}^{0\star}}
\def\W{\rm{W}^{\pm}}
\def\es{\rm{e}}
\def\ns{\rm{n}}
\def\el{\es^{-}}
\def\eb{\es^{+}}
\def\ml{\mu^{-}}
\def\mb{\mu^{+}}
\def\pr{\rm{p}}
\def\prb{\bar{\pr}}
\def\picb{\rm{pb}^{-1}}
\def\pcb{\rm{pb}}
\def\EffA{\epsilon_{\pr \es}}
\def\EffB{\epsilon_{\pr \mu}}
\def\DfA{\delta_{\EffA}}
\def\DfB{\delta_{\EffB}}
\def\ee{\eb\el}
\def\mm{\mb\ml}
\def\nn{\nu\bar{\nu}}
\def\qq{\rm{q}\bar{\rm{q}}}
\def\ll{\rm{l}^{+}\rm{l}^{-}}
\def\lij{\rm{l}_{i}\rm{l}_{j}}
\def\tt{\tau^{+}\tau^{-}}
\def\ff{f,\bar{f}}
\def\arr{\rightarrow}
\def\Bhab{\Z\,\arr \ee}
\def\Et{E_{tot}}
\def\EvP{E_{ECAL}/p}
\def\cost{|\cos\theta|}
\def\Gevc{\,{\rm GeV}/c}
\def\Gev{\,\rm GeV}
\def\lB{\ell_{B}}
\def\Nbar{n_{\bar{B}}}
\def\Spc1{\hspace{0.3in}}
\def\NIM{Nucl. Instr. and Meth.}
\newcommand {\Ionz} {$\, {\rm keV} /cm \, $}
\newcommand {\dEdx} {$ d{\rm E}/d{\rm x} \, $}


\begin{titlepage}
%
%
\begin{center}
    \large
    EUROPEAN LABORATORY FOR PARTICLE PHYSICS 
\end{center}
\begin{flushright}
    \large
    \PPEnum\\
    \Date
\end{flushright}
\bigskip\bigskip
%
%
\begin{center}
    \huge\bf\boldmath
 Search for Baryon and Lepton Number Violating
 $\Z$ Decays 
\end{center}
\bigskip
\bigskip
%
%
\begin{center}
\LARGE
 The OPAL Collaboration \\
\bigskip
\bigskip
\bigskip
\large
\end{center}
%
%
\begin{abstract}

Using data collected with the OPAL detector
at LEP, we have searched for the processes
$\ee \arr \Z \arr \pr \el , \pr \ml $ and
the charge conjugate final-states. These would
violate the conservation of the baryon-number B,
lepton-number L and the fermion-number n = (B+L). 
No evidence for such decays has been found, and the 
95~\% confidence level upper limits
on the partial widths $\Gamma(\Z \arr \pr \es )$
and $\Gamma(\Z \arr \pr \mu )$
are found to be $ 4.6 $  and $ 4.4  {\, \rm keV} $
respectively.

\end{abstract}
 
\bigskip
\bigskip
\bigskip
\bigskip\bigskip
\begin{center}
{\large (Submitted to Physics Letters B)}
\end{center}

\end{titlepage}

\begin{center}{\Large        The OPAL Collaboration
}\end{center}\bigskip
\begin{center}{
G.\thinspace Abbiendi$^{  2}$,
K.\thinspace Ackerstaff$^{  8}$,
G.\thinspace Alexander$^{ 23}$,
J.\thinspace Allison$^{ 16}$,
N.\thinspace Altekamp$^{  5}$,
K.J.\thinspace Anderson$^{  9}$,
S.\thinspace Anderson$^{ 12}$,
S.\thinspace Arcelli$^{ 17}$,
S.\thinspace Asai$^{ 24}$,
S.F.\thinspace Ashby$^{  1}$
% ...
}\end{center}\bigskip
\bigskip
$^{  1}$School of Physics and Astronomy, University of Birmingham,
Birmingham B15 2TT, UK
\newline
$^{  2}$Dipartimento di Fisica dell' Universit\`a di Bologna and INFN,
I-40126 Bologna, Italy
\newline
$^{  3}$Physikalisches Institut, Universit\"at Bonn,
D-53115 Bonn, Germany
\newline
$^{  4}$Department of Physics, University of California,
Riverside CA 92521, USA
\newline
$^{  5}$Cavendish Laboratory, Cambridge CB3 0HE, UK
\newline
$^{  6}$Ottawa-Carleton Institute for Physics,
Department of Physics, Carleton University,
Ottawa, Ontario K1S 5B6, Canada
\newline
$^{  7}$Centre for Research in Particle Physics,
Carleton University, Ottawa, Ontario K1S 5B6, Canada
\newline
$^{  8}$CERN, European Organisation for Particle Physics,
CH-1211 Geneva 23, Switzerland
\newline
$^{  9}$Enrico Fermi Institute and Department of Physics,
University of Chicago, Chicago IL 60637, USA
\newline
$^{ 10}$Fakult\"at f\"ur Physik, Albert Ludwigs Universit\"at,
D-79104 Freiburg, Germany
\newline
$^{ 11}$Physikalisches Institut, Universit\"at
Heidelberg, D-69120 Heidelberg, Germany
\newline
$^{ 12}$Indiana University, Department of Physics,
Swain Hall West 117, Bloomington IN 47405, USA
\newline
$^{ 13}$Queen Mary and Westfield College, University of London,
London E1 4NS, UK
\newline
$^{ 14}$Technische Hochschule Aachen, III Physikalisches Institut,
Sommerfeldstrasse 26-28, D-52056 Aachen, Germany
\newline
$^{ 15}$University College London, London WC1E 6BT, UK
\newline
$^{ 16}$Department of Physics, Schuster Laboratory, The University,
Manchester M13 9PL, UK
\newline
$^{ 17}$Department of Physics, University of Maryland,
College Park, MD 20742, USA
\newline
$^{ 18}$Laboratoire de Physique Nucl\'eaire, Universit\'e de Montr\'eal,
Montr\'eal, Quebec H3C 3J7, Canada
\newline
$^{ 19}$University of Oregon, Department of Physics, Eugene
OR 97403, USA
\newline
$^{ 20}$CLRC Rutherford Appleton Laboratory, Chilton,
Didcot, Oxfordshire OX11 0QX, UK
\newline
$^{ 22}$Department of Physics, Technion-Israel Institute of
Technology, Haifa 32000, Israel
\newline
$^{ 23}$Department of Physics and Astronomy, Tel Aviv University,
Tel Aviv 69978, Israel
\newline
$^{ 24}$International Centre for Elementary Particle Physics and
Department of Physics, University of Tokyo, Tokyo 113-0033, and
Kobe University, Kobe 657-8501, Japan
\newline
$^{ 25}$Institute of Physical and Environmental Sciences,
Brunel University, Uxbridge, Middlesex UB8 3PH, UK
\newline
$^{ 26}$Particle Physics Department, Weizmann Institute of Science,
Rehovot 76100, Israel
\newline
$^{ 27}$Universit\"at Hamburg/DESY, II Institut f\"ur Experimental
Physik, Notkestrasse 85, D-22607 Hamburg, Germany
\newline
$^{ 28}$University of Victoria, Department of Physics, P O Box 3055,
Victoria BC V8W 3P6, Canada
\newline
$^{ 29}$University of British Columbia, Department of Physics,
Vancouver BC V6T 1Z1, Canada
\newline
$^{ 30}$University of Alberta,  Department of Physics,
Edmonton AB T6G 2J1, Canada
\newline
$^{ 31}$Duke University, Dept of Physics,
Durham, NC 27708-0305, USA
\newline
$^{ 32}$Research Institute for Particle and Nuclear Physics,
H-1525 Budapest, P O  Box 49, Hungary
\newline
$^{ 33}$Institute of Nuclear Research,
H-4001 Debrecen, P O  Box 51, Hungary
\newline
$^{ 34}$Ludwigs-Maximilians-Universit\"at M\"unchen,
Sektion Physik, Am Coulombwall 1, D-85748 Garching, Germany
\newline
\bigskip\newline
$^{  a}$ and at TRIUMF, Vancouver, Canada V6T 2A3
\newline
$^{  b}$ and Royal Society University Research Fellow
\newline
$^{  c}$ and Institute of Nuclear Research, Debrecen, Hungary
\newline
$^{  d}$ and Department of Experimental Physics, Lajos Kossuth
University, Debrecen, Hungary
\newline
$^{  e}$ on leave of absence from the University of Freiburg
\newline

\newpage

\section{Introduction}

No symmetry principle is known in physics which may guarantee 
the conservation of baryon (B)- or lepton (L)-number.
Nevertheless, there are two important observations which are 
relevant to B and L non-conservation in a very 
profound way: namely baryogenesis~\cite{Bgenesis}, 
and the realization that B and L are not strictly conserved 
in the Standard Model of electroweak interactions~\cite{tHoof}.

In the standard hot big bang model~\cite{Bgban}, one of the
most obvious relics from the early universe is the 
baryon. But the universe, at some distance scale
around us, $\lB$, is practically 
100~\% charge asymmetric, with baryon 
number density very much exceeding that of antibaryons,
$n_{B} \gg \Nbar $ (and $n_{\el} \gg n_{\eb})$. 
The distance $\lB$ is at least of the size 
of our visible universe.  
The standard electroweak theory contains in
principle all the elements necessary 
for the generation of the baryonic asymmetry of the 
universe~\cite{Sakharov,Kuzmin}. 
These theoretical elements are 
\begin{description}
\item (i) anomalous electroweak baryon-number 
non-conservation~\cite{tHoof}, 
\item (ii) C- and CP-violation in the 
fundamental gauge and Higgs 
interactions of the quarks,  
\item (iii) deviation from thermal equilibrium, 
assuming the cosmological 
electroweak phase transition is first order~\cite{Kuzmin}, 
which occurs at temperatures
of about $\sim 100  \Gev. $\footnote{
Sakharov suggested~\cite{Sakharov}
that baryogenesis took 
place immediately after the big bang, at a temperature 
not far below the Planck scale of $10^{19}  \Gev$. This 
scenario was explicitly realized with the advent of 
the Grand Unified Theory (GUT), which predicts B and
possibly CP violations, mediated by very heavy  
gauge bosons, X and Y. The GUT baryogenesis is appealingly
simple but does not easily fit into an acceptable 
cosmology.} 
\end{description}

At low temperatures, this anomalous electroweak B 
violation is negligible~\cite{Bgenesis,sphaleron}.
For instance, the proton lifetime, determined from 
the absence of the decay 
$ \pr \arr \eb \Zs \arr \eb \eb \el$  was found 
to be larger than $5 \times 10^{32}$ years
at 90~\% confidence level~\cite{protlife}.
On the other hand, at the $\Z$ resonance with an 
equivalent temperature of 90  $\Gev$, the decays 
$\Z \arr \pr \es /\pr \mu $ may conceivably 
occur at an observable level due to the
anomalous interactions mentioned above.

In this paper, we present the results of a search for
the decays $\Z \arr \pr \el , \pr \ml $ and their charge 
conjugate final-states, $\prb \eb$ and $ \prb \mb  $, 
with the OPAL detector at LEP. 
These decays would violate B, L and the fermion number, 
n = (B+L) while 
conserving (B--L). We have searched for events 
characterized by a pair of
back-to-back charged particles,
where one is identified as a lepton and
the other as a proton.  
Although other rare decay modes of the
$\Z$ can also be handled in the same 
analysis,  they are not 
discussed here since they either violate 
conservation laws
attributable to some fundamental 
symmetries, or since stringent constraints
on their rates  have 
already been set by OPAL and other LEP 
experiments~\cite{OPAL Rares}.

\section{The OPAL detector and data sample}

A complete description of the  OPAL 
detector can be found 
in Ref.~\cite{ref:OPAL-detector}, and
only those components 
most relevant to the identification of 
high energy electrons, muons and protons 
are described briefly here. 

The central detector provides 
charged particle tracking
over 96\% of the full solid angle 
in a 0.435~T uniform magnetic field 
parallel to the beam axis~\footnote
   {The OPAL coordinate system is defined 
    so that the $z$ axis is in the
    direction of the electron beam, the $x$ 
    axis is horizontal 
    and points towards the centre of the LEP ring, and  
    $\theta$ and $\phi$
    are the polar and azimuthal angles, 
    defined relative to the
    $+z$- and $+x$-axes, respectively. 
    The radial coordinate is denoted
    as $r$.}.
It consists of a two-layer silicon microstrip 
vertex detector, a high precision drift chamber,
a large volume jet chamber, and a set of $z$ 
chambers measuring the track coordinates along 
the beam direction. The sensitive volume of 
the jet chamber is a cylinder of about 2 m in 
radius and 4 m in length. It is divided 
into 24 sectors, each equipped
with 159 sense wires and two cathode wire
planes. Up to 159 measurements of position 
and charge deposition per track are possible.
The momentum resolution of the whole 
tracking system is $\sigma_{p_t}$/$p_t$
$\approx \sqrt{0.02^2 + (0.0015 \cdot p_t)^2}$ 
where $p_t$ is the momentum component transverse 
to the beam direction. The charge measurements
were used for the calculation of the specific 
energy loss \dEdx of a track. The \dEdx resolutions 
obtained with  a number of samplings $\,N \, > 130 $ 
are $ = 3.1 \; \% $ and  $ 3.8 \; \% $ for muons in
$\Z \arr \mm $ and for minimum ionising pions within 
jets, respectively~\cite{dEdx}. A detailed description
of the main features of the jet chamber can be 
found in~\cite{Jetchamber}. 

The magnetic coil is surrounded by a time-of-flight 
counter array (TOF) which measures the time-of-flight 
for a charged particle from the interaction region with
a time resolution of about 300 ps and can be used 
for the rejection of cosmic ray muons. A lead-glass 
electromagnetic calorimeter (ECAL), located outside 
the magnet coil, covers the full azimuthal range 
with excellent hermeticity in the polar angle range 
of $\cost <0.984$. In the barrel region 
($ \cost <0.82$), the ECAL is composed of 9440 
lead-glass blocks, each 37 cm in depth. The blocks
are approximately $10 \times 10 \, cm^{2}$ 
in cross section, and the calorimeter is 
typically 24.6  radiation lengths deep. 
The energy resolution  of the barrel ECAL is
approximately $\sigma_{E}
/E \simeq 2.3 \%$ for 45 $\rm GeV$ 
electrons. The magnet return yoke is segmented 
and instrumented for hadron calorimetry (HCAL) 
and consists of barrel and endcap sections 
along with pole tip detectors that together 
cover the region $|\cos \theta |<0.99$. The
HCAL provides tracking for muons above $3 \Gev$ 
and sampling for the showers produced by 
hadrons interacting with material (a total of
$\sim$ 8 absorption lengths to the back of
HCAL in the barrel region) inside the lead 
glass of ECAL and the return yoke of HCAL. 
Four layers of muon chambers cover the outside 
of the hadron calorimeter. 

The electromagnetic calorimeters close to the beam axis 
complete the geometrical acceptance down to 24 mrad in 
$\theta$, except for the regions where a tungsten shield 
is present to protect the detectors from synchrotron 
radiation. These are lead-scintillator sandwich 
calorimeters and, at smaller angles, silicon tungsten 
calorimeters~\cite{ref:SW} located on both sides of 
the interaction point and used for luminosity  measurement.

The data used for this analysis were recorded  
at LEP during 1990 -- 1994. The integrated 
luminosity of the data sample  
is 119 $\, \picb$~\cite{Lumin} corresponding to 
5.2  million  $\Z$ decays.  

\section{Monte Carlo simulations}

The simulation of $\Z \arr \pr \el, \pr \ml $
decays and their charge conjugate final-states  
begins with 20,000 $\Z \arr \mm $ events 
that were generated at 
the $\Z$ resonance peak with 
$\sqrt{s} = 91.2 \, \Gev $ using the 
KORALZ~\cite{KorMM} Monte Carlo. 
The masses, particle types and 
the four momenta vectors of the muon pairs 
were then changed into those of  
$\pr \el, \prb \eb , \pr \ml $ or $\,\prb \mb\,$ 
pairs with equal probability and  
the necessary corrections were made
to preserve momentum and energy 
conservation. Responses of the OPAL 
detector to these particles were then 
obtained by processing the four vectors 
through the OPAL Monte Carlo simulation 
chain~\cite{OPAL7}.

The analysis procedure consists of the 
selection of a pair of collinear charged 
particles in the barrel region of the 
jet chamber, each with 
approximately the beam 
energy, followed by particle 
identification.  Major backgrounds
were anticipated from mis-identification 
of the high energy charged leptons 
in $\Z \arr \mm, \, \ee,$   and $\tt $ decays.  
Large samples of Standard Model lepton pairs 
( $ 600\,000 \; \mm, 942\,000 \; \ee $ and 
$ 447\,500 \; \tt $ ) were generated with the 
KORALZ and BABAMC~\cite{BHAMC} Monte Carlo 
programs. In addition, 1.3 million 
Standard Model $\Z$ decays into $\qq$'s,
where $ q = u,d,c,s $ and $b$ quarks,  
were also generated with the
JETSET~\cite{LUND} Monte Carlo program. 
All events are then processed through
the OPAL analysis chain~\cite{OPAL7}.

\section{Preselection} 

 In this analysis, candidate events were chosen 
from a sample of low (charge) multiplicity events. 
We demanded  that the  events 
contained two, and only two, well--measured high 
energy tracks, satisfying the following 
preselection criteria to guarantee a high
degree of redundancy in triggering and to give 
almost an 100 \% trigger efficiency for the detection 
of the signal~\cite{Lowm,Muon eff}: 

\begin{description}

\item $\bullet$  Well--measured tracks were defined by:

   The distance of closest approach to the beam axis in
the plane perpendicular to the beam axis 
must be less than 2 cm.

   The distance to the interaction point along the beam
axis, at the point of closest approach in the plane 
transverse to the beam axis, must be less than 40 cm.
 
   The $\chi^{2}/n.d.f. $ of the track fit in the $r-\phi$ 
plane must be less than 2.

   The first sense wire in the jet chamber used for 
track fitting must lie within a radius of 75 cm from 
the colliding beam axis.

   There must be at least 100 hits used for the 
ionization energy loss, \dEdx, measurement and 4 
hits in the $z$ chambers. 

    The track polar angle must satisfy $\cost < 0.7 $.  

    The track must have either its measured 
    $p_{t} \geq 15 \Gevc$,
    or its associated   
    $  E_{ECAL} > 35  \Gev$.

\item $\bullet$ To select the required event 
   topology, the two tracks were required to be: 
   
   oppositely charged and collinear within  
   $\delta\theta_{acol} < 10^{\circ}$, 

   in coincidence at the TOF counters, where 
   we demanded that: 
   \[ |  t_{1} sin \theta_{1} 
        - t_{2} sin \theta_{2}  | < 10 \; {\mathrm nsec} \] 
where $t_{1,2}$ and $\theta_{1,2}$ are the measured 
time-of-flights and polar angles of the two 
preselected particles respectively.
\end{description}

At this stage $139\,328$ events passed the event 
preselection criteria. Possible contamination
from two photon interactions were efficiently  
removed by the high $p_{t}$ and $E_{ECAL}$ cuts.
Cosmic muons passing through the interaction 
region were rejected by the coincidence requirement
on their TOF measurements. The surviving events,
consistent with the number expected from standard 
model $\Z $ decays 
to $\ee, \, \mm $ and $\tt $, constituted the major 
background to the signal.  Comparisons of the data with 
these Standard Model Monte Carlo simulated
$\Z$ decay backgrounds are shown in row 3 of 
Table~\ref{tab:tab1} for the numbers of events 
that survived preselection, and in Figures 1 and 2 
for the $\Et = E_{ECAL} + E_{HCAL}$ and \dEdx 
distributions of the charged tracks, respectively. 
The Monte Carlo simulated background reproduced 
the data very well.  

\section{\bf Particle identification}

We demanded, among the two pre-selected 
high energy particles in each event, that  
there be one identified charged lepton
and one charged track identified as a 
proton and not as a lepton.
Particle identification was achieved 
using five independent quantities, 
namely momentum ($p$), ionization energy 
loss (\dEdx),  energy deposits in the 
calorimeters, $ E_{ECAL}$ and $ E_{HCAL}$,  
and the information on matching of muon 
chamber hits to the charged track.
We defined a `tracking road' along the 
direction of each charged track in the 
OPAL calorimeters with a width: 
\[ 20 \,{\rm mrad} + 100 \,{\rm mrad}/p\;(\Gevc),
   {\mathrm \;in\;HCAL} \]  
\[100 \,{\rm mrad} + 100 \,{\rm mrad}/p\;(\Gevc),
   {\mathrm \;in\; the\,muon\,chambers.} \] 
\begin{description}
\item $\bullet$ Electron identification:

Electron candidates were defined as those tracks having 
$  E_{ECAL} > 35   \Gev$ with $\EvP > 0.8 $,  
and  no hits found in the `tracking road', 
except for the first layer 
of the HCAL. 
\item $\bullet$ Muon identification:

As muon candidates we considered those tracks
which failed the electron identification. 
Those tracks must have its measured momentum,
$p$, satisfying $ 40 \Gevc < p < 60 \Gevc $,
and have more than 4 hits along the `tracking road'. 
To distinguish muons from protons, 
we further demanded that either at least two of the 
hits on the `tracking road' be in the muon chamber,
or there be at least 4 of the hits in the HCAL with 
no more than one hit per layer, or that there 
be at least 2 of the hits in the 3 outermost 
layers of the HCAL. 
\item $\bullet$ Proton identification:
 
We considered those tracks as proton 
candidates which failed the electron
identification. The proton track must have 
$ 40 \Gevc < p < 60 \Gevc $ and have at least 
one hit found in HCAL with the 
expected total calorimeter energy associated 
with the measured track $\,\Et > 10  \Gev,$ 
but requiring that $E_{ECAL} \leq 35 \Gev $.
\item $\bullet$ Event identification:

Finally, events identified as $\ee$ or $\mm$ 
pairs were removed, while event to be identified  
as a signal must satisfy the \dEdx requirement 
defined below:
\item $\bullet$ \dEdx requirement:

The proton must have a measured 
\dEdx $<$ 9 \Ionz (see Figure 2). For 
the accompanied $\es$ or $\mu$, 
the measured \dEdx must agree with its 
expected value for the particle assignment 
and the measured momentum.
\end{description}

No events in the data sample survived 
the selection criteria, while 1.2  
background events were expected from 
the Standard Model background processes.

\section{Efficiency and systematic errors}

\begin{figure}[htbp]
\centering
\begin{tabular}{cc}
\epsfig{file=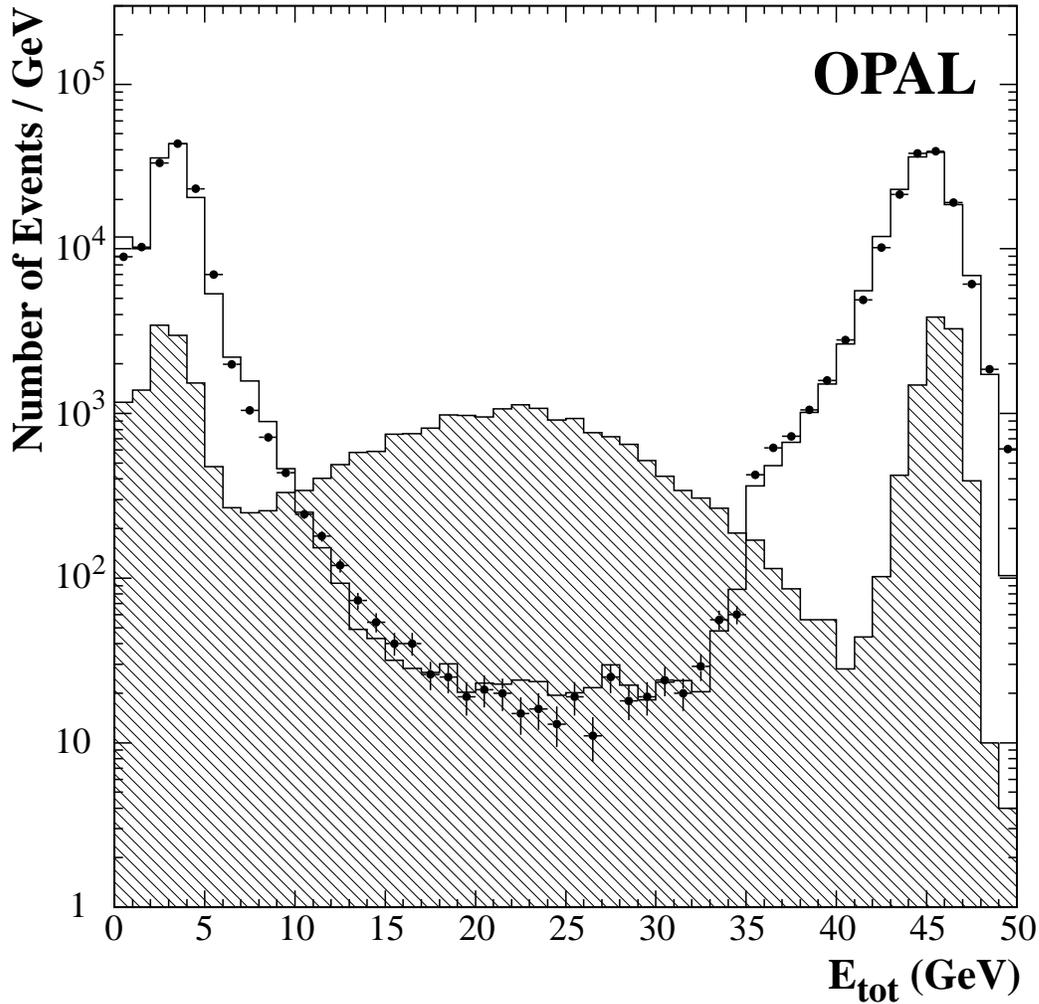,width=15cm}
\end{tabular}
\caption[]{\sl
  \protect{\parbox[t]{15cm}{

The measured calorimeter energy (data points), 
$\Et$, associated with each well measured 
charged track satisfying the 
event preselection criteria for the data.
The Monte Carlo simulated $\mm, \ee,$ and $\tt$ 
events are shown by  the open histogram
for comparison. The shaded histogram
represents the Monte Carlo simulated 
$\Z \arr \pr \es/\pr \mu $ decays
with arbitrary normalization.} } } 
\label{fig:fig1}
\end{figure}

To estimate the detection efficiency 
of the signal and to investigate 
possible bias in the selection of  
events, preselection and particle 
identification criteria were applied to
reconstructed Monte Carlo signal events.
The shaded histogram in Figure~\ref{fig:fig1},
with arbitrary normalization, shows the 
distribution of the expected total calorimeter 
energy, $\Et$, associated with each of the 
measured tracks of pre-selected 
Monte Carlo simulated signal events.
The low energy peak centered at 
3 $\Gev$ comes from the muons,
while the peak centered at the beam energy 
(45  $\Gev$) comes from the
electrons.  The broad peak in the 
middle of the spectrum comes from
the protons~\cite{OPAL7,NIMCAL}. 
No such central region broad peak  
is seen either in the data (points)
or in the Monte Carlo simulated 
background (open histogram) events as 
shown in Figure~\ref{fig:fig1}. 

\begin{figure}[htbp]
\centering
\begin{tabular}{cc}
\epsfig{file=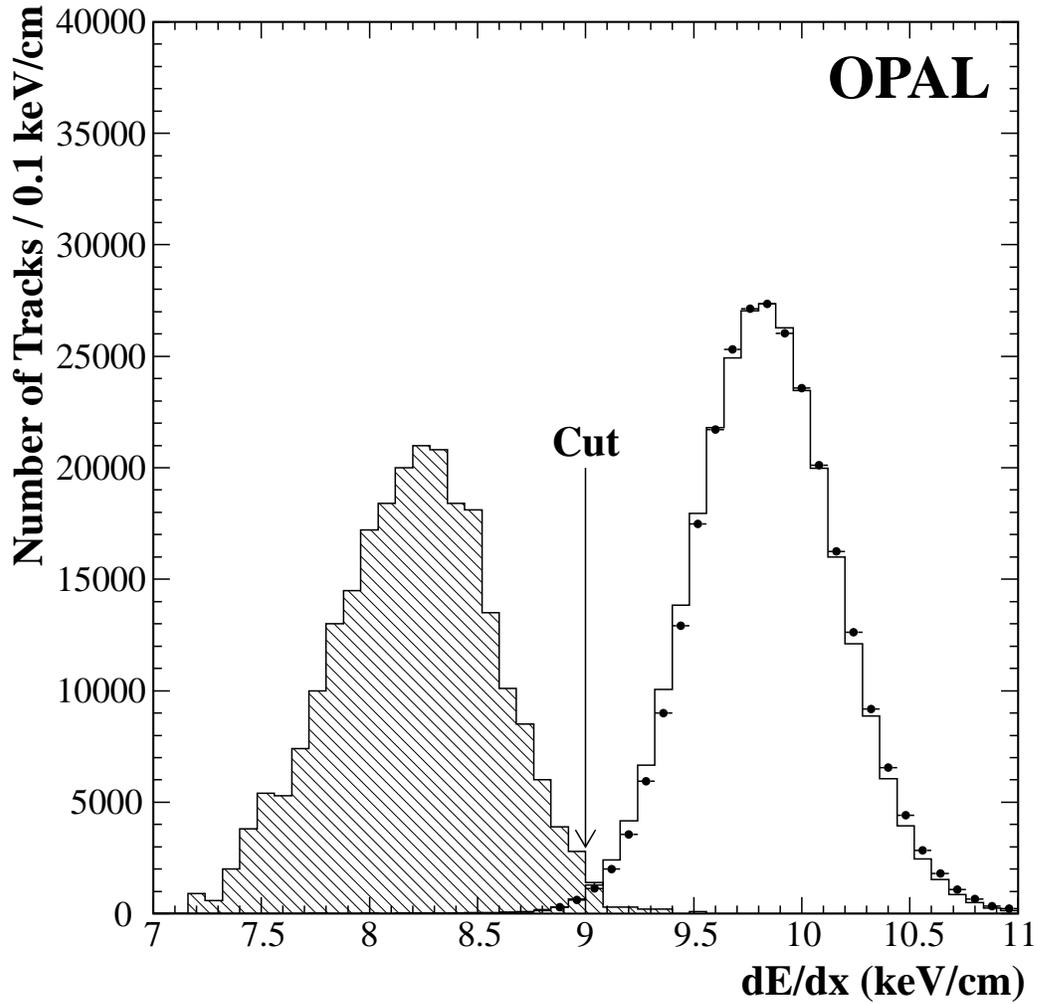,width=15cm}
\end{tabular}
\caption[]{\sl
 \protect{\parbox[t]{15cm}{ The  
distribution of specific ionization 
energy loss, \dEdx, of the well measured
charged tracks (data points) that survived the event 
preselection. The open histogram
shows the corresponding distribution for the 
Standard Model Monte Carlo simulated
background charged lepton pairs. 
The \dEdx distribution of a 45 
$\Gev $ proton which survived the same 
selection criteria is shown as the
shaded histogram with arbitrary 
normalization. } } } 
\label{fig:fig2}
\end{figure}

The preselection criteria limited the acceptance 
of the signal to the barrel region of the OPAL 
detector where the jet chamber provided a large 
number of hits for each of the pre-selected 
charged tracks for the \dEdx and momentum 
measurements. These criteria ensured a better than 5.3 \% 
resolution in momentum and a 3.5 \% resolution 
on \dEdx for a $ 45 \Gev $ charged particle. As a 
result, 3$\sigma$ separation between protons 
and muons, and 4$\sigma$ separation between 
protons and electrons at  $ 45 \, \Gev $ were 
achieved~\cite{dEdx}. The \dEdx distribution
of the pre-selected data is compared to that 
of the Monte Carlo simulated background in 
Figure~\ref{fig:fig2}. 
The shaded histogram shows the \dEdx distribution 
expected for 45 $\Gev $ protons. A \dEdx $ < 9$ 
\Ionz cut was used for the identification of protons.

Table~\ref{tab:tab1} summarizes the event flow
history. The equivalent integrated luminosities
of the Monte Carlo generated background sample
sizes are shown in row 1. Row 2 shows the numbers 
of these events, normalized to the recorded luminosity
($119\picb$), surviving the requirements of low 
(charge) multiplicity. In row 3, the numbers of 
Monte Carlo simulated events and data events 
surviving the preselection are compared. The 
agreement here is excellent. Not shown in the Table 
are the corresponding numbers for the standard 
model Monte Carlo simulated multi-hadronic decays 
of the $\Z$, because we found that none of the 
1.3 million Monte Carlo simulated $\Z \arr \qq $
events survived the preselection criteria.  In
addition, six million Standard Model $\Z$ 
decays were generated with JETSET~\cite{LUND} 
and subjected to a  preselection at the four-vector 
level. We found no event other than the Standard Model 
$\ee, \mm$ or $\tt$ pairs that survived the 
preselection criteria. In Table 1, row 4 shows 
that:
\begin{description}
\item the pre-selected data were dominated by 
      (97.4 \% ) $\ee, \mm$ pairs. 
\item after preselection, the efficiency for the
      identification of $\ee$ was found to be 
      99.9 \%, and 94.5 \% for $\mm$. 
\item after the identified $\ee$ and $\mm$ pairs were 
      removed, none of the events in the data sample 
      survived the \dEdx requirement selection, 
      while 1.2  events were expected from the 
      simulated background.
\item the detection efficiencies of the signal events 
     for $ \Z \arr \pr \mu $ and $\pr \es $, were 
     found to be: \\
     $ \EffB = 0.353 \pm  0.006(stat) \; $ and 
     $ \;  \EffA = 0.338 \pm  0.007(stat) $  
     respectively. 
\end{description}

More details on particle identification for the 
pre-selected events are shown in Table~\ref{tab:tab2}. 
The numbers of tracks surviving the preselection
criteria are shown in row 1. The numbers of electron 
and muon candidates identified in the data sample 
are compared to those of the Monte 
Carlo simulated backgrounds in rows 2 and 3. Proton 
candidates shown in row 4 are due to the 
mis-identification of ($\sim$ 0.5 \% ) the 
pre-selected charged leptons before imposing the 
final \dEdx requirement. This was anticipated because:
\begin{description}
\item $\dagger$  a high energy muon or electron can  
sometimes deposit a great deal of its energy inside 
the calorimeters due to bremsstrahlung, $\ee$ pair 
production, or deep inelastic nuclear interaction
(DIS). These processes are characterized by large 
uncertainties, small cross sections, hard spectra, 
and large energy fluctuations in the generation of
electromagnetic and hadronic showers,
\item $\dagger$ a high energy lepton can undergo DIS 
in the ECAL, where only the electromagnetic component 
of the shower is measured by the lead glass, resulting 
in $E_{ECAL} < 35 \Gev$,
\item $\dagger$ hadrons are emitted in $\tau$ decays. 
\end{description}

These effects, in particular the DIS of the $\es$'s
and $\mu$'s, made the proton identification, 
with $\Et$ cuts alone, difficult.  However, after  
removing the identified $\ee$ and $\mm$ pairs, the 
final \dEdx requirement efficiently eliminated every
one of the 1249 identified proton candidates 
in the data. The 1.2 background events, all identified 
as $\Z \arr \pr + \mu $, consisted of 
0.5 events coming from $\mm$ and 0.7 
events from $\tt$. The $\mm$ background events 
can be understood as DIS of one of the high 
energy muons, while the $\tt$ background events 
are due to mis-identification of 
$\Z \arr \tau_{1} \tau_{2} $, where 
$\tau_{1} \arr h $ and $\tau_{2} \arr \mu $. 

Allowing the \dEdx and $\Et$ cuts to vary,  
based on the analysis of the Monte Carlo 
simulated events, we estimated a 0.71 \%  
systematic uncertainty for the detection  
of the signal.

Final-state radiation (FSR) is expected to 
be different for muons and electrons or 
protons. Changing the particle code from 
$\mu$ to $\es $ in the final-state of 
$\Z \arr \mm $ events (see section 3)
may have caused an underestimation of the FSR 
correction on the electrons in the 
$\Z \arr \pr \es$ events resulting in a 
larger than expected acceptance of 
the signal. Based on a study of the $\Bhab$ 
events, we found that 0.47~\%  of the electrons 
could have a measured $E_{ECAL}$ less than 
35 $\Gev$. This value was quoted as 
the systematic uncertainty in the estimation 
of the detection efficiency due to FSR. 

Another source of systematic error came from the 
0.41~\% uncertainty in the collected 
luminosity~\cite{Muon eff}. In summary, a quadratic sum 
of all of the above mentioned errors yielded 
a 0.94~\%  systematic error on the detection 
efficiency, $\EffA$, and 0.81~\% on $\EffB$.  

\section{Results and conclusion.}

Using a data sample of 5.2 million $\Z$ decays and 
with 0 candidate event found, we obtained the 
following upper limits: 
\[ \Gamma(\Z \arr \pr \es ) =   4.6  {\, \rm keV} \]
and 
\[ \Gamma(\Z \arr \pr \mu ) =  4.4  {\, \rm keV} \]
at 95~\% confidence level~\cite{Error Bar}.

This is the first time that limits have been obtained
on the partial widths of $\Z$ decays for processes which 
simultaneously violate baryon-, lepton-, and 
fermion-number conservation. The above limits 
are comparable to the existing limits on 
lepton family number violating $\Z$ 
decays~\cite{OPAL Rares}.

\bigskip\bigskip\bigskip
\appendix
\par
{\Large\bf Acknowledgements}
\par
We particularly wish to thank the SL Division for the efficient operation
of the LEP accelerator at all energies
 and for their continuing close cooperation with
our experimental group.  We thank our colleagues from CEA, DAPNIA/SPP,
CE-Saclay for their efforts over the years on the time-of-flight and trigger
systems which we continue to use.  In addition to the support staff at our own
institutions we are pleased to acknowledge the  \\
Department of Energy, USA, \\
National Science Foundation, USA, \\
Particle Physics and Astronomy Research Council, UK, \\
Natural Sciences and Engineering Research Council, Canada, \\
Israel Science Foundation, administered by the Israel
Academy of Science and Humanities, \\
Minerva Gesellschaft, \\
Benoziyo Center for High Energy Physics,\\
Japanese Ministry of Education, Science and Culture (the
Monbusho) and a grant under the Monbusho International
Science Research Program,\\
Japanese Society for the Promotion of Science (JSPS),\\
German Israeli Bi-national Science Foundation (GIF), \\
Bundesministerium f\"ur Bildung, Wissenschaft,
Forschung und Technologie, Germany, \\
National Research Council of Canada, \\
Research Corporation, USA,\\
Hungarian Foundation for Scientific Research, OTKA T-016660, 
T023793 and OTKA F-023259.\\



\newpage 
\begin{table}[hbt]
\centering
\begin{tabular}{||c|c||c||c|c|c||c||c|c||} 
        \hline\hline
    &  &  Data  & 
     \multicolumn{6} {c||}  {\em Monte Carlo} \\ \cline{4-9}
    &  & Collected in & \multicolumn{4} {c||}  
       {\em Background}&
      \multicolumn{2} {c||}  {\em signal}  \\ \cline{4-9}
    &  & 1990-1994  & ($\ee$) & ($\mm$) &  ($\tt$)& total&
       ($\pr,\es$) &($\pr,\mu $) \\ \hline
    &  & & & & & & & \\
 1. & Sample size & 119 &  252 & 463 & 345 &  
    &  10,000 & 10,000  \\ 
    & used in  & ($\picb$) & ($\picb$) & ($\picb$) & ($\picb$) & 
    & (events) & (events)  \\ 
    & analysis  & & & & & & 100 \% & 100 \% \\ \hline\hline
    &  Multiplicity   & & & & & & & \\
 2.  & and Status & 1,896,853 & 440,027 & 148,053 &  141,117  
          & 729,197 & 92 \% & 93 \%  \\
    & requirements & & & & & &  &  \\ \hline
    &              & & & & & &  & \\
 3. & preselection & 139,328 &  73,855 &  65,436 & 149 & 
         139,440 & 48 \% & 50 \%  \\
    &              & &  & & & &  & \\ \hline 
    &              & &  & & & &  &  \\
    &  &  135,679 &  73,797 & 61,807 & 84 & 135,688 & 
          13 \% &  14 \% \\
    & Particle 
    &  \multicolumn{7} {c||} {\em (identified 
       \hspace{.5in} as \hspace{.5in}
      $ \Z \arr \ee, $ \hspace {.5in} or
       \hspace {.5in} $ \mm $)}  \\ \cline{3-9}  
 4. &  &    &    &      &     &  &   & \\ 
    &  identification  & 0  & 0  & 0.5  & 0.7 & 
                    1.2 & 33.8 \%  &  35.3 \%  \\ 
    &  & 
    \multicolumn{7} {c||} {\em (\hspace {.5in} identified 
     \hspace{.5in} as  \hspace{1.0in} signal\hspace {.5in})}   
     \\ \hline \hline 
\end{tabular} 
\caption[]{\sl
\protect{\parbox[t]{15cm}{
Comparison of event flows for data and Monte 
Carlo simulated events. Numbers shown in rows 
2, 3 and 4 represent events surviving the selection 
criteria. The Monte Carlo background event 
numbers were normalized to the luminosity of 
the data. } } }
\label{tab:tab1}
\end{table}

\newpage 
\begin{table}[hbt]
\centering
\begin{tabular}{||c||c||c|c|c||c||} 
        \hline\hline
     &  Data  & 
     \multicolumn{4} {c||}  {\em Monte Carlo} \\ \cline{3-6}
      & Collected in & \multicolumn{4} {c||}  
       {\em Background} \\ \cline{3-6}
      & 1990-1994  & ($\ee$) & ($\mm$) &  ($\tt$)& total
       \\ \hline
      & & & & &  \\
  preselection & 278,656 & 147,710  & 130,872 & 298 
    &             278,880     \\
    &              & & & &  \\ \hline \hline 
    &              & & & &  \\
  Electron &  147,527 &  147,467 & 93   & 151   & 
                 147,711    \\
     candidates  & &  & & & \\ \hline \hline
                 & &  & & & \\
   Muon    &  124,996  &       3 &  125,142  &  41 & 
                 125,186    \\
   candidates & &  & &  &  \\ \hline \hline 
        & & & & &  \\    
 Proton   & 1,249 & 53  &  914 & 78 & 1,045 \\  
 candidates &  & &  & & \\ \hline \hline 
            &  & &  & & \\
 Proton    & 0   & 0  &  0.5  & 0.7 & 1.2   \\ 
 identified   &     &    & 
      \multicolumn{3} {c||} {\em as \hspace{.5in} 
       $\Z \arr \mu + \pr $ } 
            \\ \hline \hline 
\end{tabular} 
\caption[]{\sl
\protect{\parbox[t]{15cm}{
Comparison of track flows for data and Monte 
Carlo simulated events during  
particle identification. Numbers shown in the
Table give the number of tracks surviving  
the selection criteria. } } }
\label{tab:tab2} 
\end{table}

\newpage

\newpage

\end{document}